\documentclass{aastex6}

\begin{document}


\title{SNaX - A Database of Supernova X-ray Lightcurves}


\author{Mathias Ross\altaffilmark{1} and Vikram V. Dwarkadas \altaffilmark{2}}
\affil{Astronomy and Astrophysics, University of Chicago, 5640 S Ellis Ave, ERC 569, Chicago, IL 60637}
\altaffiltext{1}{Mathias\_Ross@msn.com} 
\altaffiltext{2}{vikram@oddjob.uchicago.edu}



\begin{abstract}
We present the Supernova X-ray Database (SNaX), a compilation of the
X-ray data from young supernovae (SNe). The database includes the
X-ray flux and luminosity of young SNe, days to years after
outburst. The original goal and intent were to present a database of
Type IIn SNe. After having accomplished this we are slowly expanding
it to include all SNe for which published data are available. The
interface allows one to search for SNe using various criteria, plot
all or selected data-points, and download both the data and the
plot. The plotting facility allows for significant
customization. There is also a facility for the user to submit data
that can be directly incorporated into the database. We include an
option to fit the decay of any given SN lightcurve with a
power-law. The database includes a conversion of most datapoints to a
common 0.3-8 keV band so that SN lightcurves may be directly compared
with each other. A mailing list has been set up to disseminate
information about the database.  We outline the structure and function
of the database, describe its various features and outline the plans
for future expansion.

\end{abstract}




\section{Introduction} 
\label{sec:intro}
Young Supernovae (SNe) are objects that are observable over almost the
entire wavelength range, from radio to gamma-rays\footnote{No young SN
  has been observed in gamma-rays, and only upper limits exist
  \citep{hesslmc}. Nonetheless it is expected that that they do emit
  at $\gamma$-ray wavelengths, but are beyond the detection limits of
  current telescopes \citep{kdb95, vt09, vvd13, mrdt14}}, days to
years after the SN explosion.  Observations at optical wavelengths
have been the norm in the last 400 years since the invention of the
telescope.  Observations at X-ray and radio wavelengths were generally
not possible until the modern era; advances in technology and
instrument building in the last 50 years, coupled with the launch of
space-based observatories, have led to a significant increase in X-ray
observations of supernova remnants (SNRs) and young SNe, measuring
X-ray emission days, years, or even decades after explosion. The
increased sensitivity of X-ray observatories like {\it Chandra} allows
us to delineate a young SN from its neighbors with arc-second
precision, and provides us with high resolution spectra of the
same. For SNe in not so crowded fields {\it XMM-Newton} is an equally
good option, while the {\it Swift} telescope has become the go-to
instrument for initial detection. More recently, {\it NuStar} has
provided a reasonably high resolution instrument to detect and obtain
spectra of objects in the range of 6-79 keV for the first time ever,
opening up new windows for observations of young SNe.

With the introduction of modern-day X-ray observatories the number of
young X-ray SNe detected is steadily growing. The first SN to be
detected in X-rays was SN 1980K, using the {\it Einstein} observatory
\citep{canizaresetal82}. This was followed by the X-ray detection of
SN 1987A, initially with {\it Ginga}, and then with every other X-ray
satellite that has operated after the time of explosion. Due to its
proximity, SN 1987A {\textbf has been extremely well-observed in
  X-rays}, and has the most detailed and complete X-ray light
curve. SN 1986J, SN 1978K and SN 1993J round out the first five
\citep{Schlegel95} detected X-ray SNe. The pace has certainly picked
up since then, especially with the launch of {\it Chandra} and {\it
  XMM-Newton} in 1999, followed by {\it Swift}. In 2003, \citet{il03}
counted 15 X-ray SNe. The number had grown to 25 by 2006
\citep{schlegel06}, and 32 by 2007 \citep{immler07}. In 2012,
\citet{dg12} collected together known lightcurves of most SNe, and
listed 42 distinct X-ray detected SNe.  This number has now grown to
over 60. We note that as of the writing of this paper, all the known
X-ray detected SNe are core-collapse SNe arising from massive stars.
No X-ray detection of a Type Ia SN has as yet been reported.

As the rate of discovery and the number of data-points continues to
grow it becomes more essential, and more difficult, to look up
available information about detected X-ray SNe. Now is the time to
create a well-thought-out database -- while the amount of data is
manageable, albeit growing quickly, rather than once the dataset
becomes too large to easily assemble. There exist well maintained
databases at other wavelengths, especially in the optical, such as
WISeREP \citep{wiserep}. However, till now no central repository has
existed for X-ray observations. Recently, \citet{opencatalog} have
created a repository to host all available observations of SNe at all
wavelengths, especially optical, X-ray and radio. This is a catalog
that seeks to pull in existing data from other catalogs into one
central database. While it is a commendable and ambitious endeavor, it
does depend on the availability of the data already in other catalogs,
or the willingness of the user community to contribute data. It has no
special emphasis on the X-rays, and does not attempt to homogenize the
X-ray data and present results in a common waveband for comparison as
is done herein.

In this paper we announce the creation of a catalog specifically to
hold X-ray observations of young SNe, the
\underline{S}uper\underline{N}ov\underline{a} \underline{X}-ray
Database, abbreviated as SNaX.  SNaX can be found at the URL: {\em
  http://kronos.uchicago.edu/snax}. SNaX is a moderated, easy-to-use
interactive database of X-ray observations of SNe. In \S 2 we outline
the rationale for such a database and the science that can be carried
out using SN X-ray lightcurves. In \S 3 we describe the database, its
structure and contents, and the back end. In \S 4 we illustrate the
graphical interface, the layout and access to the data. \S 5
distinguishes the database from similar endeavors and discusses our
views for the future of the database.

\section{Rationale for the SNaX database} 
\label{sec:lightcurve}

\subsection{ What can one learn from X-ray lightcurves and Spectra of SNe?}
Thermal X-ray emission, including thermal bremsstrahlung and line
emission, depends on the square of the density of the emitting
material. In a SN the emitting material is generally the post-shocked
material, whose density is 4 times that of the pre-shock medium for a
strong adiabatic shock. Thus, thermal X-ray emission is immediately a
tracer of the density of the material into which the shock is
expanding. The evolution of the X-ray lightcurve can be used to
understand the density structure of the material, which can be related
to the mass-loss from the progenitor star, and therefore to the SN
progenitor. More sophisticated (and somewhat approximate) analysis can
help to understand the medium structure even in the case of weak,
radiative and/or reverse shocks \citep{cf94, flc96,
  dg12,chandraetal09, chandraetal12, drrb16}. If the emission is
non-thermal, the mechanism is generally inverse Compton or synchrotron
emission, as postulated for Type IIP \citep{cfn06} and Type Ib/c
\citep{cf06} SNe. The X-ray lightcurves can also be used to decipher
the density structure in these cases \citep{sayanetal13, sayanetal16}.

These relationships mean that X-ray lightcurves can allow us to (1)
distinguish between adiabatic and radiative shocks from SNe in certain
circumstances, and (2) distinguish between expansion into a steady or
non-steady wind which may arise from the SN progenitor\footnote{As of
  this writing, X-ray emission has been detected only from
  core-collapse SNe, which arise from massive stars that have strong
  winds that modify their circumstellar environment.  Upper limits are
  available for Type Ia SNe \citep{ri12, marguttietal14}}. In some
cases the winds may create circumstellar wind-blown bubbles, further
complicating the environment \citep{vvd05}, and having significant
implications for the X-ray emission. Lightcurves further (3) allow us
to derive the slope of the density profile of the medium into which
the SN is expanding, (4) reveal details about the SN kinematics,
dynamics and evolution, when coupled with spectral information and
other multi-wavelength data, and (5) provide valuable information
regarding elemental abundances, both in the shocked circumstellar
medium formed by the SN progenitor and in the SN ejecta.

\subsection{Goals of the SNaX database} 
It is clear that there is a need to compile the X-ray fluxes of young
SNe, which has led to the creation of the SNaX database. The main
goals for the creation of the SNaX database are the following:

\begin{enumerate}
\item To have a moderated database which will (eventually) present all
  available data on X-ray SNe.
\item To provide an easily accessible and searchable user interface to
  rapidly find the X-ray data on any given SN.
\item To provide a template that allows users to easily submit data to the
  database, so that it can be stored and retrieved.
\item To provide a graphical user interface enabling the user to plot
  the data and download plots.
\item To provide X-ray fluxes and luminosities in a standard waveband
  so that all the data can be compared. One of the problems outlined
  in \citet{dg12} was that it was very difficult to compare
  lightcurves of various SNe, or indeed even create a lightcurve for a
  given SN, because different authors provide fluxes and/or
  luminosities in different wavebands. We have found fluxes for
  different SNe quoted in various bands, including: 0.2-2 keV, 0.4-2
  keV, 0.3-8 keV, 0.4-2.4 keV, 0.5-4 keV, 0.5-8 keV, 2-8 keV, 3-8 keV,
  and 2.4-10 keV.  SNe discovered with {\it Swift} have their flux
  generally quoted in the 0.2-10 keV range. This makes it extremely
  difficult to make quantitative comparisons between the data from
  various SNe.  Thus, although the first step is to present all the
  data, {\em an essential component} is to provide X-ray flux values
  in a given common band. We have chosen the 0.3-8 keV band for this
  purpose, which is relevant to {\it Chandra}, {\it XMM-Newton} and
  {\it Swift} data. We urge authors who study X-ray SNe to adopt the
  same philosophy and provide fluxes in the 0.3-8 keV band.
\item To intentionally maintain a moderated database. We feel that
  this is necessary in order to maintain data integrity, evaluate
  submissions, and provide fluxes in a standard waveband so that they
  can be easily compared.
\end{enumerate}

\section{Structure of the Database}
The data are stored in a {\sc MySQL} relational database.  A web
interface has been developed using a combination of {\sc php} and {\sc
  javascript}, with graphing functionality provided by the open source
{\sc flot} graphing library (which is in turn built upon the {\sc
  jquery} javascript library), available at
http://www.flotcharts.org/. The use of {\sc MySQL} allows for rapid
data retrieval, and scales smoothly with an increase in the number of
X-ray SNe.  Only the plotting and analysis features make extensive use
of javascript, ensuring that a large amount of computing power is not
required by users looking up data.

\subsection{Database Tables}
The core of the data is stored in three MySQL tables.  At the
`highest' level is a table containing data pertaining to the supernova
itself, such as distance, explosion date, and coordinates.  Connected
to the supernova table is a table containing information determining
the source of the observation: instrument used, dates covered in the
observation, exposure time, etc.  A third table contains measurements
taken from each observation: count rates, flux,errors, and the energy
range over which the flux was measured.  Each entry also has an
associated citation, stored as a bibcode to easily generate a link to
the NASA ADS database.

There are two further tables that are connected to the three main
tables, and are used for storing data that needs arbitrary numbers of
rows.  The first is used to store the IDs associated with the
observations; for example, a list of Chandra obsIDs.  The second
stores model information in fields like column density $N_{H}$ and
temperature; if the model takes into account a variety of abundances
they are also stored here.

\subsection{Calculated Data}
The database uses the stored data to calculate the age of the SN at
the time of observation, and its luminosity.  Although flux data
stored for a single SN may originate from a variety of papers, SNaX
only records a single value for object overview information, such as
the object's distance.  This means that a given X-ray luminosity may
differ from that listed in the publication in which the flux
measurement is listed, should an alternate value for distance be used.
We have tried not to be judgemental in choosing which distance value
is used, but if different papers give the luminosity for an object
using different distances, we need to make a choice to use one of the
given distances, thus leading to a change in some of the luminosities.
Generally, values recorded by SNaX coincide with those used by the
more recent publication, with the rationale that authors are aware of
previous measurements; similar concerns arise with age calculations
based upon an explosion date.  If a publication provides only the flux
values we have adopted a distance from the {\sc SIMBAD}
\citep{simbaddb} or {\sc NED} databases for use in luminosity
calculations.

\section{Data Layout and Access}

Currently our goal is to include all available {\it published} data on
X-ray SNe in the database, by which we mean data having an ADS
entry. We also include conversions to the 0.3-8 keV energy range,
which we calculate when it is not provided. We do not include any data
that exists only in poster form. In some cases, although the data is
published, quantitative information--such as a table of numbers--is
not available even though a lightcurve is shown; in such cases it is
of course impossible for us to include this in our database. Having
already come across a few such publications, we urge authors to
provide tables of numbers along with graphs showing the flux or
luminosity lightcurve.

The database is funded by a NASA grant to study Type IIn SNe, and its
original intent was to present all X-ray data on Type IIn SNe only. As
such, incorporating all the available data on Type IIn SNe into the
database has been our first priority. Having mostly accomplished this,
and having created a working framework and interface, our ongoing goal
is to expand the database to include data on all X-ray SNe, and
furthermore to provide fluxes in the 0.3-8 keV band whenever possible.
We realise this will take sustained time and effort, and are carefully
working on this task. We are also hopeful that users will submit
information on other SN types to the database, helping us in our task
of making this an extremely useful resource. Some data on other SNe is
already available, for example SN 1987A \citep{franketal16}.

\begin{figure*}[htbp]
\includegraphics[width=\textwidth]{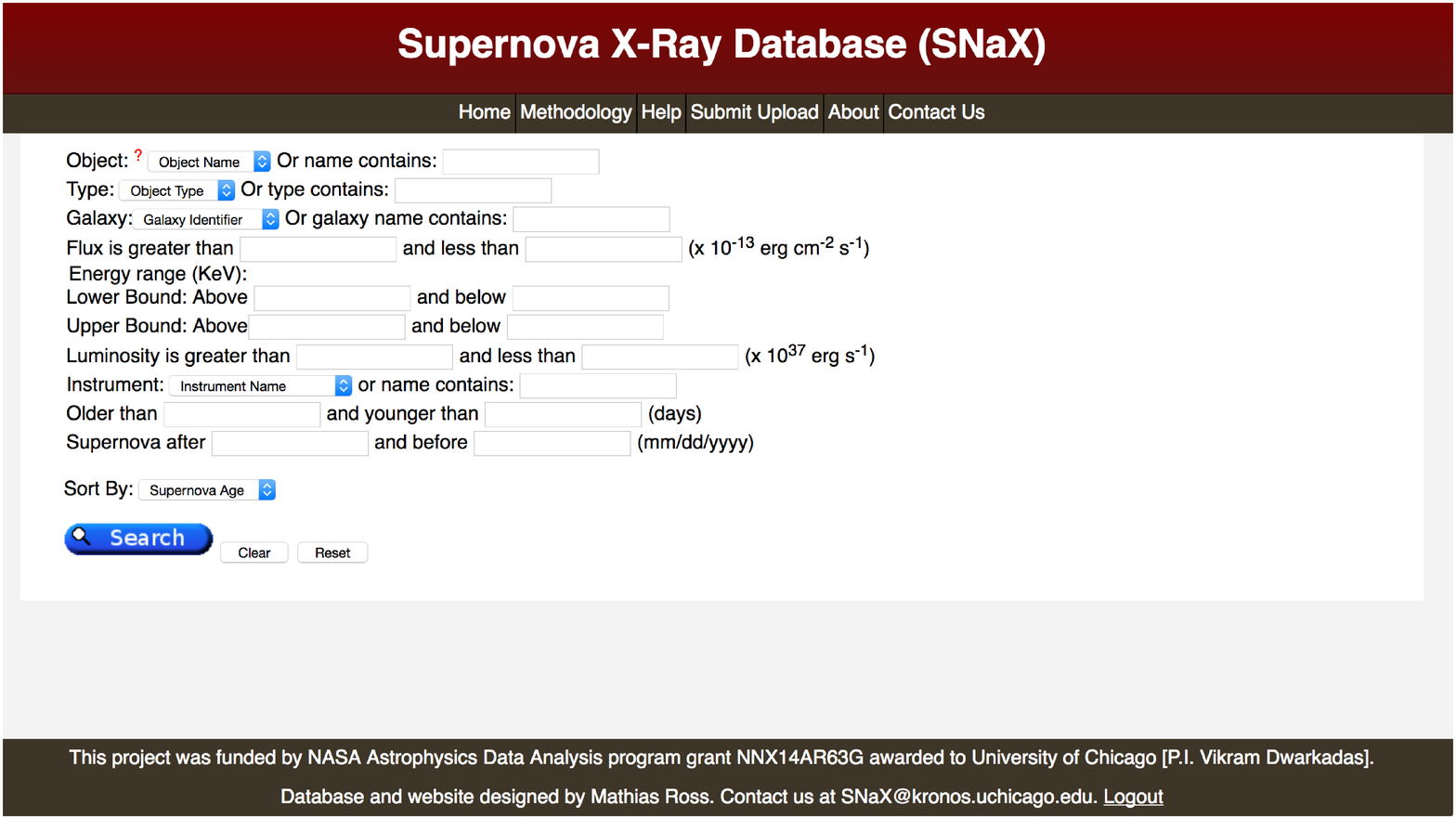}
\caption{The home page that greets the user when they enter the
  database. The powerful search interface allows searches
  by name, host galaxy, type, luminosity bounds, explosion date,
  instrument etc.  }
\label{fig:db_home}
\end{figure*}

\subsection{Homepage} 
\begin{figure*}[htbp]
\includegraphics[width=\textwidth]{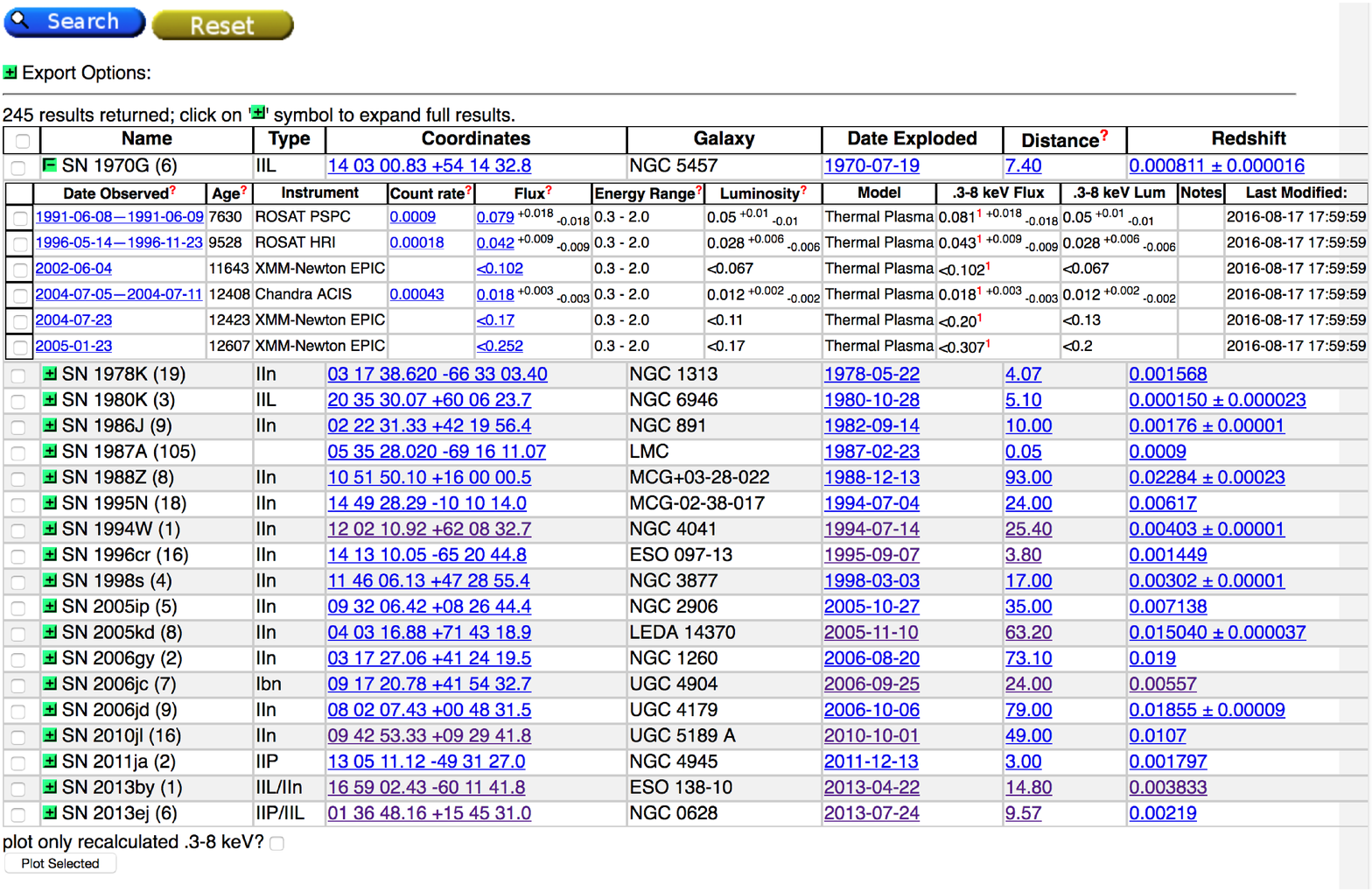}
\caption{A list of SNe is displayed according to the search
  criteria. The list can be sorted by any of the listed parameters. If
  no search criteria are specified, pressing the blue `Search' button
  brings up all the SNe in the database. Pressing the `$+$' next to a
  SN name brings up the various datasets associated with the
  particular object. Clicking on quantities having hyperlinks
  generally brings up the link to the citation from which the
  particular parameter was derived.  }
\label{fig:db_datapoints}
\end{figure*}

Upon entering the database, the homepage initially offers the user a
choice of selecting the SN via several options (Figure
\ref{fig:db_home}): specifying the name of the individual SN,
searching by SN type, specifying the luminosity within a given range
(upper and lower bounds can be individually specified), searching for
a given host galaxy, explosion date, age, satellite or instrument
used, and other criteria. This opens up a list of SNe (Figure
\ref{fig:db_datapoints}). Once a SN or group of SNe is selected, a
single click expands each SN dataset to show all the datapoints
available for a particular SN. If multiple models are available giving
different luminosities, or when different groups provide alternate
models for the same dataset, we have included all the available models
and flux parameters. Data are given in hypertext throughout the
website, as a link to the ADS entry for the paper from which the
information was taken.  It is also possible to see the bibcode of the
relevant citation by hovering the mouse over the relevant information.

\subsection{Standardized Waveband for Comparison} 
A major goal of this database is to not only make data on X-ray SNe
easily accessible, but also provide standardized data sets that can be
easily compared. We have chosen to provide fluxes in the standardized
range of 0.3-8 keV for each and every SN in the database. This band
was chosen because it represents the usable overlapping range of the 3
current operational X-ray satellites {\it Swift, Chandra and
  XMM-Newton}, which account for most X-ray observations of young SNe
taken at the present time.  In order to do this expeditiously, we have
written a program that takes the data stored in the database, runs it
through a locally stored version of HEASARC
PIMMS\footnote{https://heasarc.gsfc.nasa.gov/cgi-bin/Tools/w3pimms/w3pimms.pl}
\citep{pimms} and converts the energy band to 0.3-8 keV. This is done
for those SNe where sufficient data is available to allow for the
models to be identified and PIMMS conversion to take place. It is
unable to take into account elemental abundances; when SNaX displays a
conversion from PIMMs, it assumes solar abundances, even if the
original measurement had various elements with abundances fit
according to the chosen model. Our conversion is also unable to take
into account multiple model components. These limitations are inherent
to PIMMS.

In some cases we have downloaded and fitted the data, and recalculated
the flux in the 0.3-8 keV band. These instances are marked, and our
goal is to eventually do this for several datapoints, especially for
those that have high-resolution {\it Chandra} and {XMM-Newton}
data. We note that this may lead to an inconsistency in that the 0.3-8
keV flux may be computed from a different model fit than was used for
the fluxes listed at other wavebands. While we endeavor to check that
our model gives about the same flux as listed in the database at other
wavelengths, this is not always possible, especially in cases where
full original model details are not provided. We urge all submitters
to provide data in the 0.3-8 keV band in addition to whatever other
bands they may have computed the flux. We also include an option to
plot only the 0.3-8 keV flux.

\subsection{Plotting} 
\begin{figure*}[htbp]
\includegraphics[width=\textwidth]{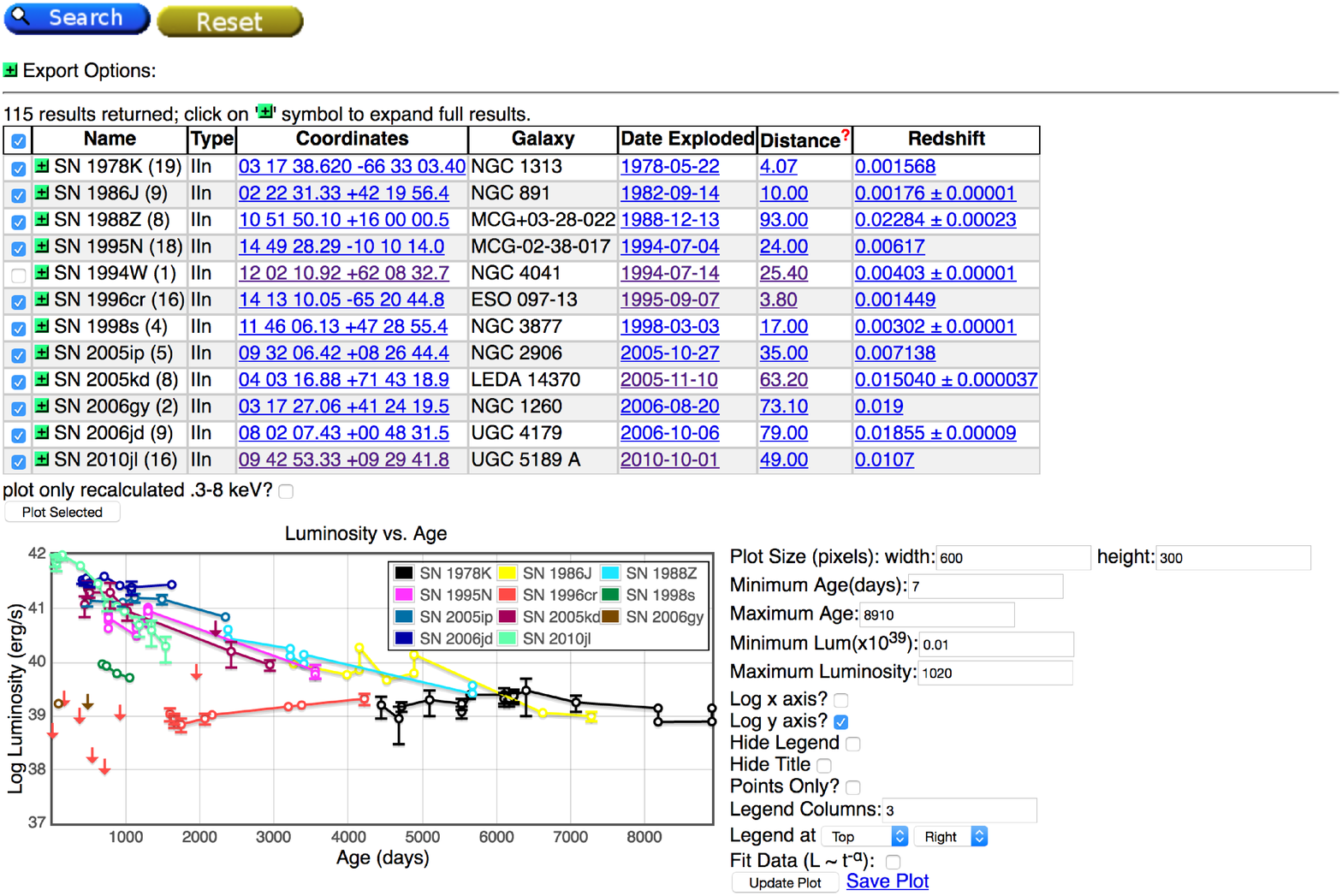}
\caption{Plotting the SN lightcurves in the database. Individual SNe
  can be selected to be plotted, and if necessary individual points
  from each of those can be selected. Upper limits are also
  displayed. The plot can be fully customized, including the nature of
  the axes, the minimum and maximum for each axis, the legend
  orientation and number of columns.  }
\label{fig:db_plot}
\end{figure*}

The interface allows the user to easily plot all the data for a given
SN or multiple SNe. Alternatively, specific datapoints can be selected
from the available list for every individual SN. There is also an
option to plot the recalculated 0.3-8 keV data to allow proper
comparison between the lightcurves of various SNe. The plot itself is
fully customizable (Figure \ref{fig:db_plot}), down to the size, axis
range, axis type (linear or logarithmic), plot titles etc. The legend
can be located at any of the four corners, and distributed over a
given number of desired columns so as not to overlap any data. Both
the plot and the data itself are available for download by the user.

\subsection{Lightcurve Fitting} 
\begin{figure*}[htbp]
\includegraphics[width=\textwidth]{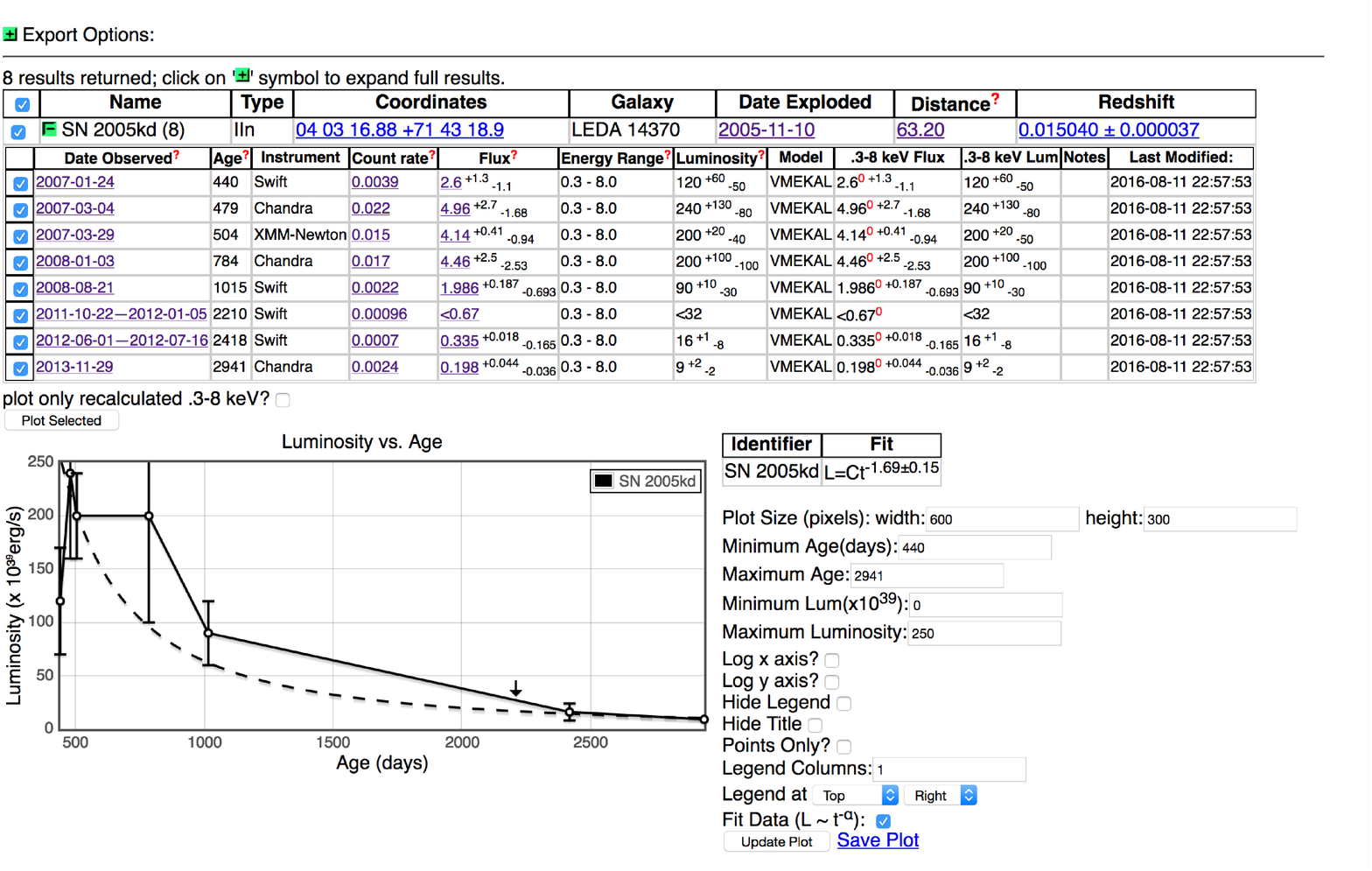}
\caption{An individual object, SN 2005kd, is selected and all points
  plotted. The option to fit a power-law to the data is selected. The
  best fit is displayed as a dashed line on the plot, and the fit
  parameters listed at the top right of the image. }
\label{fig:db_fit}
\end{figure*}

The decay of the X-ray lightcurve can provide information about the
density structure of the ambient medium in which the SN is expanding
(see \S \ref{sec:lightcurve}). As first postulated by
\citet{chevalier82}, and further elucidated in \citet{flc96} and
\citet{dg12}, if the self-similar solution for the expansion of the SN
shocks with time \citep{cf94} can be assumed, the X-ray luminosity
evolves as a power-law with time.  The SNaX database incorporates an
option to compute the slope of the lightcurve, assuming that the
luminosity (`L') varies with time (`t') as a power-law ($L \propto
t^{m}$). In order to calculate the power law index, SNaX runs a linear
least squares regression on the site, using the method outlined in
{\it Numerical Recipes in C} \citep{numrecbook}. After taking the
natural logarithm of both sides, a linear fit is used to find the
slope `m' and intercept `b' of the expression ${\rm ln}(L) = m ({\rm
  ln}\, t) + b$. The final fit is displayed on the screen at the top
right of the plot (Figure \ref{fig:db_fit}). When the flux values have
known uncertainties $\sigma_{i}$ this regression minimizes the value
of $\sum_{i=1}^{N} (\frac{y_{i} - b -mx_{i}}{\sigma_{i}})^2$ (for N
data points).  The fitting algorithm assumes symmetric Gaussian error;
when the error is asymmetric SNaX is programmed to take the lower
error bound to be the value of $\sigma_{i}$. Upper limits are ignored
when computing the fit.

\subsection{Data Uploads and Downloads}
Data can be quickly viewed on the site, and lightcurves plotted for
easy comparisons. If desired the data may be easily downloaded for
off-line use in a comma-separated-variable (csv) format (which can be
opened by almost any spreadsheet software, such as Excel) or
tab-separated-variable (tsv) format which can be opened using a text
editor.

We have included a facility for the user to upload data to the
database. Using a web form, it is possible to submit data
pre-formatted into a csv file. An excel spreadsheet-like template
provides the user with a sample of the layout of the data.  The
submitted file is parsed by the host, and a preview of the data as it
would appear in the database is displayed as long as javascript is
being utilized. If the user accepts the preview and submits, an
automatic email reply is generated. The data are temporarily stored in
a duplicate form of the database structure, awaiting review from a
moderator.  If the data have been correctly parsed by the system they
are transferred to the main database for access; if possible,
conversion of flux data to the 0.3-8 keV band will happen at this
time.

We have also included a basic API, to provide users with the ability
to integrate custom applications with the SNaX database.  Any database
search query that can be generated using the primary html interface
can be duplicated through this API, which provides data formatted
using JSON.  A more detailed description of the functionality provided
is available at www.kronos.uchicago.edu/snax/apiDocumentation, and the
interface itself is at www.kronos.uchicago.edu/snax/urlQuery.

\subsection{Requirements and Testing}

SNaX has been tested on at least the following systems: Windows (7,
8.1 and 10), MacOS (10.6.8, 10.10.5), and Linux (Ubuntu 12.04, Ubuntu
16.04, CentOS 7). In all cases it has been found to work properly, and
perform as desired, using Firefox and Google Chrome browsers. Safari
on newer systems works fine, although on an older Mac using 10.6.8,
Safari 5.1 was found to have problems. Performance was as desired with
the new Edge browser on Windows, and it also worked with Internet
Explorer 10 on Windows 7.  In general the latest versions of Firefox,
Google Chrome and Windows Edge browsers should work on all systems,
and meet performance expectations.

The database is accessible on mobile devices, although it is not
necessarily responsive. The graph size will not adapt to the size of
the browser window, and left to right scrolling may be necessary to
read listed data. It has been tested on devices including an ASUS T100
tablet running Windows (with Firefox), an IPad Mini (Chrome, Safari
browsers), an Amazon Kindle (Silk Browser), and an Android phone
running Android 5.0.2 (Chrome, Opera Browsers). In all cases the
performance was satisfactory.  On devices that are unable to utilize
javascript it is still possible to view and export data, although
plotting and analysis options may be unavailable.

Our first implementation of the database was opened up to a few users
in different countries in March 2016, who accessed a beta version and
provided comments. The testers were generally satisfied with the user
interface and contents, but had comments on available or additional
features, search criteria, or modifications that could be made. These
comments were used to make several improvements to the database,
including host galaxy search and lightcurve fitting. A help section
providing further usage details has been added, and will be
continually updated. We think the database is now working robustly and
can be opened up to astronomers worldwide. The developers welcome
feedback on the structure, interface and implementation aspects of the
database, that would allow us to continually improve its usefulness.

\section{Discussion and Future Prospects}
We have presented in this paper a new database, SNaX, that lists the
X-ray data from young SNe. The database can be searched in various
ways, allows for plotting of selected data or the complete dataset,
and includes a functionality to upload data. All data can be
downloaded if desired. Over the next few years we anticipate adding
almost all available data on X-ray SNe to the database. We are hopeful
that users will contribute a substantial amount of data as well to
make this the go-to site for astronomers needing X-ray data on
supernovae.

This database differs from other available databases, such as the Open
SN Catalog, in some important ways:

\begin{itemize}
\item It allows for comparison of the lightcurves. One can plot the
  lightcurves of all SNe, or any specific ones as needed, on a single
  plot, allowing for easy comparison. To our knowledge this is not
  possible with other available catalogs at present.
\item We aim to (eventually) provide data on all SNe in a single X-ray
  band (0.3-8 keV) so as to enable proper comparison. In the past this
  was impossible, as can be seen from the published datapoints, which
  are given in a multitude of wavebands.
\item The moderators (currently this paper's authors, although that
  could change in future) are actively looking through the published
  astrophysics literature for listed X-ray fluxes of SNe, and adding
  them to the database.
\item Where available, we also store and provide model information,
  which is very useful for reproducing the analysis, or recalculating
  the flux in various bands.
\item Submissions from other astronomers are welcome and
  appreciated. A template is provided to ensure that data are
  submitted in the correct format. The download is inspected by the
  moderators before being added to the database.
\item The moderators are continually downloading and fitting data, and
  providing fluxes for those SNe whose data is available but not
  published.
\item SNaX allows the user to fit the lightcurve of one or more SNe
  with the press of a single button, plot the fit and list the
  power-law index.
\item It is moderated to ensure catalog integrity, prevent duplicates
  and ensure valid data uploads.
\end{itemize}

We have created a mailing list to post announcements and enable
discussion of the database. Information and subscription details can
be found at: https://lists.uchicago.edu/web/info/snax-users

One of our future goals is to include the reduced X-ray spectra of
young SNe within the database. For this to happen, astronomers working
in this area need to provide their spectral files in {\sc FITS} format
along with the data in published papers. There is not currently any
provision to do so, but we hope that this changes over time, and that
astronomers will start to provide the reduced spectra. In the meantime
the database moderators are tasked with developing an interface to
adequately deal with spectra in the FITS format.  This is one of the
major aspects to be explored in future.

We believe that the SNaX database can become a go-to resource for
X-ray data on young supernovae. We hope to include all SNe as the
number of SNe detected in X-rays increases, and cover all SN types. We
encourage astronomers to use it and provide constructive feedback to
the developers\footnote{The developers and moderators can be directly
  contacted via the email address listed in the database,
  snax@kronos.uchicago.edu} so that we may continue to improve the
infrastructure, contents and delivery. We do request that all users
please cite this paper, which enables a counting of aggregate users
and thus establishes the usefulness of the database, in addition to
citing all sources of the data that they download.

\acknowledgements We acknowledge comments and suggestions from the
anonymous referee that helped to improve this paper in many ways. This
work is supported by a NASA Astrophysics Data Analayis program grant
\# NNX14AR63G awarded to PI V.~Dwarkadas at the University of
Chicago. We thanks Dr F.~Bauer, Dr. F.~Sutaria and Dr. R.~Chevalier
for comments on an earlier version of the database, and Dr. K.~Frank
for providing a table of SN 1987A data. This research has made use of
NASA's Astrophysics Data System; the SIMBAD database operated at CDS,
Strasbourg, France; and the NASA/IPAC Extragalactic Database (NED)
which is operated by the Jet Propulsion Laboratory, California
Institute of Technology, under contract with the National Aeronautics
and Space Administration.

\bibliographystyle{aasjournal} \bibliography{references}

\label{lastpage}



\end{document}